\documentclass[a4paper]{jpconf}
\usepackage{graphicx}
\usepackage{amsfonts}

\usepackage{amsmath}
\usepackage{amsthm}
\usepackage{amssymb}

\newcommand\be{\begin{equation}}
\newcommand\ee{\end{equation}}
\newcommand\bea{\begin{eqnarray}}
\newcommand\eea{\end{eqnarray}}

\newcommand\minus\backslash

\theoremstyle{definition}

\theoremstyle{remark}

\begin{document}
\title{Superintegrable systems with spin induced by coalgebra symmetry}

\author{D. Riglioni $^{1*}$, O. Gingras $^{1 \dagger}$,  P. Winternitz $^{1,2 \ddagger}$}
 
\address{$^1$ Centre de Recherches Mat\'ematiques, Universit\'e de Montr\'eal, \\ CP 6128, Succ. Centre Ville,  Montr\'eal, Quebec H3C 3J7, Canada \\
$^2$ D\'epartement de Math\'ematiques et Statistique, Universit\'e de Montr\'eal, \\ CP 6128, Succ. Centre Ville,  Montr\'eal, Quebec H3C 3J7, Canada}

\ead{$^*$riglioni@crm.umontreal.ca, $^\dagger$ gingras.ol@gmail.com, $^\ddagger$wintern@crm.umontreal.ca}

\begin{abstract} 
A method for deriving superintegrable Hamiltonians with a spin orbital interaction is presented. The method is applied to obtain a new superintegrable system in Euclidean space $\mathbb{E}_3$ with the following properties. It describes a rotationally invariant interaction between a particle of spin $\frac{1}{2}$ and one of spin 0. Its Hamiltonian commutes with total angular momentum $\vec{\mathcal{J}}$ and with additional vector integrals of motion $\vec{X}$ , $\vec{Y}$ with components that are  third order differential operators. The integrals of motion form a polynomial algebra under commutation.  The system is exactly solvable (in terms of Laguerre polynomials) and the bound state energy levels are degenerate and described by a Balmer type formula. When the spin orbital potential is switched off the system reduces to a hydrogen atom. 
\end{abstract}


\section{Introduction}
The purpose of this article is twofold. First, we present a method for generating superintegrable systems with a spin-orbital interaction in three-dimensional Euclidean space $\mathbb{E}_3$ from superintegrable scalar systems in $\mathbb{E}_2$. The method starts with a Hamiltonian of the form 

\begin{equation}
H^{(2)} = \frac{1}{2} (p_1^2 + p_2^2 ) + V(\rho) , \quad \rho=\sqrt{x_1^2 +x_2^2}
\end{equation}

and uses coalgebra symmetry to generate systems of the form

\begin{equation}
\label{eqi2}
H^{(3)} = \frac{1}{2} (p_1^2 + p_2^2 + p_3^2 ) + V_0(r) + V_1(r) (\vec{\sigma} , \vec{L})
\end{equation}

where 

\begin{equation}
p_k = -i\hbar \frac{\partial}{\partial x_k}, \quad L_k = \epsilon_{k a b } x_a p_b , \quad r= \sqrt{x_1^2 + x_2^2 + x_3^2}
\end{equation}

and $\sigma_k$ are the Pauli matrices

\begin{equation}
\label{paulisigma}
\sigma_1 = \left(
\begin{array}{cc}
0 & 1 \\
1 & 0 
\end{array}
\right) , \quad \sigma_2 = 
\left(
\begin{array}{cc}
0 & -i \\
i & 0 
\end{array}
\right), \quad \sigma_3 = \left(
\begin{array}{cc}
1 & 0 \\
0 & -1 
\end{array}
\right)
\end{equation}

\noindent Secondly, we use this coalgebra method to derive a maximally superintegrable system with the Hamiltonian 

\begin{equation}
\label{eqa1.4}
H = \frac{-\hbar^2}{2} \nabla^2 + \frac{2 \gamma}{r^2} \vec{S} \cdot \vec{L} - \frac{\alpha}{r} + \frac{\hbar^2 \gamma (\gamma + 1)}{2 r^2}, \quad \vec{S} = \frac{\hbar}{2} \vec{\sigma}
\end{equation}

\noindent where $\alpha$ and $\gamma$ are arbitrary constants. This Hamiltonian is integrable because it is spherically symmetric and hence the angular momentum 

\begin{equation}
\vec{\mathcal{J}} = \vec{L} +  \vec{S}
\end{equation}

\noindent is an integral of the motion. We shall show below that the system is also superintegrable. The additional integrals of motion are the components of a vector that is in general a third order polynomial in the momenta (a third order Hermitian operator). The Hamiltonian (\ref{eqa1.4}) can be viewed as describing the Coulomb interaction of a particle with spin $\frac{1}{2}$  with another of spin $0$. We recall here that a superintegrable system is one that has more integrals of motion than degrees of freedom. The best known superintegrable systems are the only two spherically symmetrical ones  (for a particle with spin zero in a scalar potential field, or two particles with spin zero interacting via a potential depending only on the distance between them ) namely the  Kepler-Coulomb system and the isotropic harmonic oscillator \cite{1} \cite{2} \cite{3}. For a recent review of scalar  superintegrability see \cite{4}. Superintegrable systems in quantum mechanics are of physical interest for several reasons. First of all, all maximally superintegrable systems presently known are exactly solvable. This means that their energy levels can be calculated algebraically and their wave functions are polynomials in the appropriate variables, multiplied by an overall gauge function. It has been conjectured \cite{5} that all maximally superintegrable systems are exactly solvable. The second point of interest is that the energy levels of superintegrable systems are degenerate and indeed this degeneracy is explained by the non Abelian algebra of the integrals of motion. This algebra is in itself interesting. In the simplest cases the integrals  of motion form a finite dimensional Lie algebra under commutation. In other cases the Lie algebra can be a Kac-Moody algebra \cite{6}, or a finitely generated polynomial algebra (quadratic, cubic or higher)\cite{7}.  

 Earlier articles on superintegrability for particles with spin concerned fermions in external fields, usually magnetic ones (\cite{8} - \cite{15}), Schr\"odinger-Pauli equations involving a spin-orbital interaction (\cite{12},\cite{14},\cite{16},\cite{17},\cite{18}), or relativistic equations for fermions \cite{19},\cite{20}. 
\newline The connection between coalgebra symmetry and superintegrability was established by Ballesteros et al. (for a review with references to the original articles see \cite{21}). One of the present authors introduced a method based on coalgebra symmetry to derive higher order superintegrable systems from two dimensional second order ones \cite{22}, in particular the TTW and PW systems \cite{23},\cite{24},\cite{25}. The articles \cite{17} and \cite{18} were devoted to a systematic search for superintegrable systems of the form (\ref{eqi2}) with integrals of motion that are matrix polynomials of order 1 and 2 in the momenta, respectively. One of the systems obtained in \cite{18} is of the form (\ref{eqa1.4})  however with $\gamma = \frac{1}{2}$. One of the vector integrals of motion found there can be interpreted as a Laplace-Runge-Lenz vector $\vec{X}$. It is of second order in the momentum $\vec{p}$. Together with the total angular momentum $\vec{\mathcal{J}}$ these integrals generate an $o(4)$ algebra (for bound states). We shall show that (\ref{eqa1.4}) is superintegrable for all values of $\gamma$. For $\gamma = \frac{1}{2}$ or $\gamma = 1$ the vector $\vec{X}$, $\vec{Y}$ can be reduced to second order ones, for $\gamma \neq 0, \frac{1}{2}, 1$ we have third order integrals $\vec{X}$, $\vec{Y}$. The algebra generated by $\vec{\mathcal{J}}$ and $\vec{X}$ and $\vec{Y}$ is
in general a polynomial algebra that in the specific cases $\gamma = 0,\frac{1}{2},1$ turns out to be a Lie algebra isomorphic to $o(4)$. The system (\ref{eqa1.4}) for $\gamma = \frac{1}{2}$ was investigated and solved by Nikitin \cite{12},\cite{13}, together with other spin dependent systems with an $o(4)$ Fock symmetry. 

In this paper we present a series of transformations that preserve superintegrability and introduce a spin-orbit interaction into an originally scalar superintegrable Hamiltonian. 
\newline The specific example we start with is the two dimensional spinless hydrogen atom, the final result is the spin $\frac{1}{2}$ Hamiltonian (\ref{eqa1.4}) with arbitrary $\gamma$.

\section{A two-dimensional superintegrable Coulomb system with a velocity dependent potential }      

Let us start from the quantum Coulomb system in $\mathbb{E}_2$ in polar coordinates

\begin{equation}
\label{eq1}
\hat{H}_0^{(2)} = - \frac{\hbar^2}{2} \left( \partial_r^2 + \frac{1}{r} \partial_r + \frac{1}{r^2} \partial_\phi^2 \right) - \frac{\alpha}{r} 
\end{equation}

\noindent Its symmetry algebra is generated by the angular momentum $\hat{L}^2$ and the Laplace-Runge-Lenz vector $\vec{\mathcal{R}}$ with:

\begin{equation}
\label{eq2}
\hat{L}_0^{(2)} = -i \hbar \partial_\phi \quad, \quad \mathcal{R}_{1,0} = \frac{1}{2} \left( \hat{p}_2 \hat{L}_0^{(2)} + \hat{L}_0^{(2)} \hat{p}_2 \right) - \frac{\alpha x_1}{r} \quad, \quad \mathcal{R}_{2,0} = -\frac{1}{2} \left( \hat{p}_1 \hat{L}_0^{(2)} + \hat{L}_0^{(2)} \hat{p}_1 \right) - \frac{\alpha x_2}{r}.
\end{equation}

Let us use a gauge transformation $U(1) = e^{i \gamma \phi}$ to transform the superintegrable system (\ref{eq1})  (\ref{eq2}) into a velocity dependent Coulomb system that is also superintegrable, namely 

\begin{equation}
\label{eq3}
e^{-i \gamma \phi} \hat{H}_0^{(2)} e^{i \gamma \phi} \equiv \hat{H}^{(2)} = - \frac{\hbar^2}{2} \left( \partial_r^2 + \frac{1}{r} \partial_r + \frac{1}{r^2} \partial_\phi^2 \right) + \frac{\hbar \gamma}{r^2} (-i \hbar \partial_\phi) + \frac{\hbar^2 \gamma^2}{2 r^2} - \frac{\alpha}{r} 
\end{equation}

\begin{eqnarray}
\label{eq4}
e^{- i \gamma \phi} \hat{L}_0^{(2)} e^{i \gamma \phi} & \equiv & \hat{L}^{(2)}= \hbar (-i \partial_\phi + \gamma)  \nonumber \\
e^{- i \gamma \phi} \hat{\mathcal{R}}_{1,0}^{(2)} e^{i \gamma \phi} & \equiv & \hat{\mathcal{R}}_1^{(2)} = \frac{1}{2} \left( \hat{p}_2 \hat{L}^{(2)} + \hat{L}^{(2)} \hat{p}_2 \right) + \frac{i \hbar^2 \gamma x_2}{2 r^2} + \frac{\hbar \gamma x_1 }{r^2} \hat{L}^{2} - \frac{\alpha x_1}{r}\\
e^{- i \gamma \phi} \hat{\mathcal{R}}_{2,0}^{(2)} e^{i \gamma \phi}& \equiv & \hat{\mathcal{R}}_2^{(2)} =  -\frac{1}{2} \left( \hat{p}_1 \hat{L}^{(2)} + \hat{L}^{(2)} \hat{p}_1 \right) - \frac{i \hbar^2 \gamma x_1}{2 r^2} + \frac{\hbar \gamma x_2 }{r^2} \hat{L}^{2} - \frac{\alpha x_2}{r}. \nonumber
\end{eqnarray}

The momentum (or velocity) dependent Hamiltonian (\ref{eq3}) can be rewritten as

\begin{equation}
\label{eq5}
\hat{H}^{(2)} = - \frac{\hbar^2}{2} \left( \partial_r^2 + \frac{1}{r} \partial_r \right)  - \frac{\alpha}{r} + \frac{1}{2 r^2} \left( \hat{L}^{(2)} \right)^2 .
\end{equation}

We separate variables in the usual manner putting 

\begin{equation}
\label{eq6}
\hat{H}^{(2)} \psi = E \psi \quad , \quad \hat{L}^{(2)} \psi = \hbar (m + \gamma) \psi \quad , \quad \psi = R_{E m} (r) e^{i m \phi} 
\end{equation}

and obtain the radial equation

\begin{eqnarray}
\label{eq7}
\hat{H}_m^{(2)} R_{E,m} (r) & = & E R_{E, m} (r) \\
\hat{H}_m^{(2)} & = & - \frac{\hbar^2}{2} \left( \partial_r^2 + \frac{1}{r} \partial_r \right)  - \frac{\alpha}{r} + \frac{\hbar^2}{2 r^2} \left( m + \gamma \right)^2 . \nonumber
\end{eqnarray}

Let us introduce the "radial" ladder operators 

\begin{eqnarray}
\hat{a}_m^\dagger & = & \frac{1}{\sqrt{2}} \left( -i \hbar \partial_r + \frac{i \alpha}{\hbar (m + \frac{1}{2} +\gamma)} - i \hbar \frac{m + \gamma +1}{r}  \right) \\
\hat{a}_m & = & \frac{1}{\sqrt{2}} \left( -i \hbar \partial_r - \frac{i \alpha}{\hbar (m + \frac{1}{2} +\gamma)} + i \hbar \frac{m + \gamma}{r}  \right) 
\end{eqnarray}

satisfying

\begin{equation}
\hat{a}^\dagger_m \hat{a}_m= \hat{H}_{m}^{(2)} + \frac{\alpha^2}{2 \hbar^2 (m + \frac{1}{2} + \gamma)^2} .
\end{equation}

The Hamiltonian (\ref{eq7})  satisfies 

\begin{eqnarray}
\hat{a}_m \hat{H}_m^{(2)} = \hat{H}_{m+1}^{(2)} \hat{a}_m \\
\hat{H}_m^{(2)} \hat{a}_m^\dagger = \hat{a}_m^\dagger \hat{H}_{m+1}^{(2)}
\end{eqnarray}

\noindent and can hence be viewed as being shape invariant. We shall sometimes replace the eigenvalue $\hbar(m + \gamma)$ by the differential operator $\hat{L}^{(2)}$ (see [my] ). To do this without introducing negative powers of differential operators we redefine the ladder operators as

\begin{eqnarray}
\label{eq2.19}
\hat{A}^\dagger(\hat{L}) = i \hat{a}^\dagger(\hat{L}) (\hat{L}^{(2)} + \frac{\hbar}{2}) \!\! \! \! &=& \! \! \!\! \frac{1}{\sqrt{2}} \left( \hbar (\hat{L}^{(2)} + \frac{\hbar}{2})\partial_r - \alpha + \frac{(\hat{L}^{(2)} + \frac{\hbar}{2}) (\hat{L}^{(2)} + \hbar)}{r}  \right) \\
\label{eq2.20}
\hat{A}(\hat{L}) = -i \hat{a}(\hat{L}) (\hat{L}^{(2)} + \frac{\hbar}{2}) \! \! \! \! &=& \!\! \! \! \frac{1}{\sqrt{2}} \left( -\hbar (\hat{L}^{(2)} + \frac{\hbar}{2})\partial_r - \alpha + \frac{(\hat{L}^{(2)} + \frac{\hbar}{2}) (\hat{L}^{(2)} }{r}  \right)
\end{eqnarray} 

These operators satisfy 

\begin{eqnarray}
\hat{A} (\hat{L}^{(2)})  \hat{H}^{(2)} (\hat{L}^{(2)}) = \hat{H}^{(2)} (\hat{L}^{(2)} + \hbar) \hat{A} (\hat{L}^{(2)})  \\
\hat{H}^{(2)} (\hat{L}^{(2)}) \hat{A}^\dagger (\hat{L}^{(2)}) = \hat{A}^\dagger (\hat{L}^{(2)}) \hat{H}^{(2)} (\hat{L}^{(2)} + \hbar)     
\end{eqnarray}

We also define the ladder operators for the angular momentum, namely 

\begin{eqnarray}
\label{eq17}
L^+ = e^{i \phi} \quad ; \quad \hat{L}^{(2)} L^+ = L^+ ( \hat{L}^{(2)} + \hbar ) \\
\label{eq18}
L^- = e^{-i \phi} \quad ; \quad \hat{L}^{(2)} L^- = L^- ( \hat{L}^{(2)} - \hbar )
\end{eqnarray}

\noindent The $\mathbb{E}_2$ Laplace-Runge-Lenz vector (\ref{eq4}) is now expressed as 

\begin{eqnarray}
\label{quadraticrungeX}
X \equiv \vec{\mathcal{R}}_1^{(2)} = \frac{1}{\sqrt{2}} \left( L^+ \hat{A}(\hat{L}) + \hat{A}^\dagger \hat{L}^- \ \right) \\
\label{quadraticrungeY}
Y \equiv \vec{\mathcal{R}}_2^{(2)} = \frac{1}{i \sqrt{2}} \left( L^+ \hat{A}(\hat{L}) - \hat{A}^\dagger \hat{L}^- \ \right) 
\end{eqnarray} 

Finally, the superintegrable system that we are going to apply the co-algebra symmetry to is given by the Hamiltonian (\ref{eq5}) and the integrals  (\ref{eq4}). They form a symmetry algebra satisfying

\begin{eqnarray}
\left[ \hat{H}^{(2)} ,\hat{\mathcal{R}}_1^{(2)} \right] = & \left[ \hat{H}^{(2)} , \hat{L}^{(2)} \right] & =  \left[ \hat{H}^{(2)} ,\hat{\mathcal{R}}_2^{(2)} \right] = 0 \\
\left[ \hat{\mathcal{R}}_1^{(2)}, \hat{L}^{(2)} \right] \!\!\!\!\!\!\!\!\!\!\!\! &=&  \!\!\!\!\!\!\!\!\!\!\!\! -i \hbar \hat{\mathcal{R}}_2^{(2)} \\
\left[ \hat{\mathcal{R}}_2^{(2)}, \hat{L}^{(2)} \right] \!\!\!\!\!\!\!\!\!\!\!\! &=& \!\!\!\!\!\!\!\!\!\!\!\! i \hbar \hat{\mathcal{R}}_1^{(2)} \\
\left[ \hat{\mathcal{R}}_1^{(2)},  \hat{\mathcal{R}}_2^{(2)} \right] \!\!\!\!\!\!\!\!\!\!\!\! &=& \!\!\!\!\!\!\!\!\!\!\!\! -2 i \hbar \hat{L}^{(2)} \hat{H} (\hat{L}) .
\end{eqnarray}

\noindent For $H(\hat{L}) = E < 0$ this algebra is isomorphic to $o(3)$. 

\section{Coalgebra symmetry and N dimensional extensions of spherically symmetric systems}

Let us first introduce an abstract $\mathfrak{sl}(2,\mathbb{R})$ Lie algebra with basis $\hat{J}_3, \hat{J}_+, \hat{J}_-$ and commutation relations

\begin{equation}
\label{eqsl2}
[\hat{J}_3 , \hat{J}_+] = 2 i \hbar \hat{J}_+ \quad ; \quad [\hat{J}_3 , \hat{J}_-] = -2 i \hbar \hat{J}_- \quad ; \quad [\hat{J}_- , \hat{J}_+] = 4 i \hbar \hat{J}_3  
\end{equation}

we equip this algebra with a trivial coproduct $\Delta$ defined by 

\begin{equation}
\Delta(1) = 1 \otimes 1 \quad ; \quad \Delta(J_i) = J_i \otimes 1 + 1 \otimes J_i  \quad , \quad i=+,-,3
\end{equation}

The action of the coproduct defines an isomorphism for the algebra

\begin{equation}
[\Delta(\hat{J}_3) , \Delta (\hat{J}_+) ] = 2 i \hbar  \Delta (\hat{J}_+ )  \! \!\! \! \quad ; \quad \! \! \! \! [  \Delta (\hat{J}_3 )  ,  \Delta (\hat{J}_-) ] = -2 i \hbar  \Delta (\hat{J}_- ) \! \! \! \! \quad ; \quad \! \!\! \! [  \Delta (\hat{J}_-) ,  \Delta (\hat{J}_+) ] = 4 i \hbar  \Delta (\hat{J}_3 )  
\end{equation}

The coproduct can be used to generate multivariable realizations of a Lie algebra from single variable ones. We start from a single variable representation of $\mathfrak{sl}(2)$, namely 

\begin{equation}
D(\hat{J}_+) = -\hbar^2 \partial_x^2 \quad ; \quad D(\hat{J}_-) = x^2 \quad ; \quad D(\hat{J}_3) = -i \hbar (x \partial_x + \frac{1}{2})
\end{equation}  

Let us define an iteration of the coproduct $\hat{J}_i^{(n)} = \Delta(\Delta(...\Delta(\hat{J}_i)))$. The n-variable realization of $\mathfrak{sl}(2)$ is given by

\begin{equation}
\label{eq29}
D(\hat{J}_+^{(n)}) = -\hbar^2 \sum_{i=1}^n \partial_{x_i}^2 \quad ; \quad D(\hat{J}_-^{(n)}) =  \sum_{i=1}^n x_i^2 \quad ; \quad D(\hat{J}_3^{(n)}) =-i \frac{n \hbar}{2} -i \hbar \sum_{i=1}^n (x_i \partial_{x_i})
\end{equation}

The coalgebra symmetry ensures that the algebra elements  $\hat{J}_i^{(n)}, (i=+,-,3)$ commute with a set of operators obtained by a $k$ fold $(2 \leq k \leq n)$  "left" or "right" application of the coproduct $\Delta$  :

\begin{eqnarray}
\left[ \hat{J}_{+,-,3} ,\hat{\mathcal{C}}\right] &=& 0 , \quad \hat{\mathcal{C}} = \frac{1}{2} \left( \hat{J}_- \hat{J}_+ + \hat{J}_+ \hat{J}_-  \right) - \hat{J}_3^2 \\
\hat{\mathcal{C}}^{(i)} &=& \Delta(\mathcal{C})^{i} \otimes \underbrace{1 \otimes ... 1}_{n-i} \\
\hat{\mathcal{C}}_{(i)} &=& \underbrace{1 \otimes ... 1}_{n-i}\otimes \Delta(\mathcal{C})^{i} 
\end{eqnarray}

\noindent The operators $\mathcal{C}^{(k)} $ and $ \mathcal{C}_{(k)} $ can be viewed as  Casimir operators of $o(k)$ subalgebras of the $o(n)$ algebra of angular momentum. Since $\mathcal{C}^{(n)} = \mathcal{C}_{(n)} $ we have $2n-3$ Casimirs $\mathcal{C}^{(2)}, ... , \mathcal{C}^{(n)}, \mathcal{C}_{(2)},..,\mathcal{C}_{(n-1)} $. Let us now return to the realization of $\mathfrak{sl}(2,\mathbb{R})$  (\ref{eqsl2}). We rewrite the ladder operators $A$ , $A^\dagger$ and the Hamiltonian $H$ in terms of the generators $J_k$ of $\mathfrak{sl}(2,\mathbb{R})$  and its Casimir $\hat{\mathcal{C}}$. We shall use the $n$ variable realization $J_k^{(n)}$ 

\begin{eqnarray}
\label{Lalg}
(\hat{L}^{(n)})^2 \equiv \hat{\mathcal{C}}^{(n)} + \hbar^2 \!\!\!\! \quad & &   \\
\label{eq3.40}
\hat{A}^{(n)} (\hat{J}_-^{(n)} , \hat{J}_3^{(n)} , \hat{L}^{(n)}) \!\!\!\! &\equiv& \!\!\!\!\! \frac{1}{\sqrt{2}} \!\! \left(\!\! -i \frac{(\hat{L}^{(n)} + \hbar \gamma + \frac{\hbar}{2})}{\sqrt{J_-^{(n)}}} \hat{J}_3^{(n)} \!\! - \alpha \! + \! \frac{(\hat{L}^{(n)} + \hbar \gamma + \frac{\hbar}{2})(\hat{L}^{(n)} + \hbar \gamma + \hbar)}{\sqrt{J_-^{(n)}}}  \!\! \right) \\
\label{eq3.41}
\hat{A}^{\dagger (n)} (\hat{J}_-^{(n)} , \hat{J}_3^{(n)} , \hat{L}^{(n)}) \!\!\!\! &\equiv& \!\!\!\! \frac{1}{\sqrt{2}} \left( i \frac{(\hat{L}^{(n)} + \hbar \gamma + \frac{\hbar}{2})}{\sqrt{J_-^{(n)}}} \hat{J}_3^{(n)} - \alpha + \frac{(\hat{L}^{(n)} + \hbar \gamma + \frac{\hbar}{2})(\hat{L}^{(n)} + \hbar \gamma )}{\sqrt{J_-^{(n)}}}  \right) \\
\label{eq3.42}
\hat{H} (\hat{J}_-^{(n)} , \hat{J}_3^{n)} , \hat{L}^{(n)}) \!\!\!\! &\equiv& \!\!\!\! \frac{1}{2} \left( \frac{1}{J_-^{(n)}} (\hat{J}_3^{(n)} + i \hbar)^2 + \frac{(\hat{L} + \hbar \gamma )^2}{\hat{J}_-^{(n)}} \right) - \frac{\alpha}{\sqrt{\hat{J}_-^{(n)}}}
\end{eqnarray}

\noindent For $n=2$ this coincides with the formulas of Section 2. Thus $\hat{L}^{(2)}$ in (\ref{Lalg}) coincides with $\hat{L}_0^{(2)}$ in (\ref{eq2}) , $\hat{A}^{(n)}$ and $\hat{A}^{\dagger (n)}$ of (\ref{eq3.40}, \ref{eq3.41}) coincide with ( \ref{eq2.19} , \ref{eq2.20} ).The Hamiltonian (\ref{eq3.42}) reduces to (\ref{eq5}) for $n=2$. Equation (\ref{Lalg}) defines $(\hat{L}^{(n)})^2$ rather than $\hat{L}^{(n)}$ itself. For $n=2$ this is no problem since we have $\hat{L}_0^{(2)} = -i \hbar \partial_\phi$ which is the square root of $\mathcal{C}^{(2)} + \hbar^2 = -\hbar^2 \partial_\phi^2$. In Section 4 we shall need $\hat{L}^{(3)} = \sqrt{\mathcal{C}^{(3)} + \hbar^2}$ and obtain it as a linear operator satisfying $[J_k , \hat{L}^{(n)}] = 0 (k= +, - ,3)$ For any analytical function $F(z)$ we have also the commutation relations

\begin{equation}
[J_k, F(\hat{L^{(n)}})] = 0, \quad [J_3, F(J_{\pm})] = \pm 2 i \hbar J_\pm F'(J_{\pm})
\end{equation}

\noindent  and hence

\begin{eqnarray}
\label{eq37}
\hat{A}^{(n)} \hat{H}(\hat{J}_-^{(n)} , \hat{J}_3^{(n)} , \hat{L}^{(n)}) & = &  \hat{H}(\hat{J}_-^{(n)} , \hat{J}_3^{(n)} , \hat{L}^{(n)} + \hbar) \hat{A}^{(n)} \\
\label{eq38}
\hat{H}(\hat{J}_-^{(n)} , \hat{J}_3^{(n)} , \hat{L}^{(n)}) \hat{A}^{\dagger (n)} & = & \hat{A}^{\dagger (n)}  \hat{H}(\hat{J}_-^{(n)} , \hat{J}_3^{(n)} , \hat{L}^{(n)} + \hbar)
\end{eqnarray}

We wish to apply the coalgebra formalism to objects that are not necessarily purely radial, such as for instance the "angular" ladder operators $\hat{L}^+$ , $\hat{L}^-$. To achieve this we extend the algebra $sl(2,\mathbb{R})$ to a semidirect product with the Heisenberg algebra. More specifically we take the n-variable realization of $\mathfrak{sl}(2)$ (\ref{eq29}), obtained from the single variable one via the iterated co-product and extend it by the Heisenberg algebra $\mathbb{H}_n$ in $\mathbb{E}_n$.We put

\begin{eqnarray}
 \left[\hat{J}_+^{(n)} , x_k \right] = - 2 i \hbar \hat{p}_k \quad & ; & \quad \left[\hat{J}_+^{(n)} , \hat{p}_k \right] = 0 \\
 \left[\hat{J}_-^{(n)} , x_k \right] = 0  \!\! \quad  \quad \quad \quad & ; & \quad \left[\hat{J}_-^{(n)} , \hat{p}_k \right] = 2 i \hbar x_k \\
 \left[\hat{J}_3^{(n)} , x_k \right] = - i \hbar x_k \quad \quad & ; & \quad \left[\hat{J}_3^{(n)} , \hat{p}_k \right] = i \hbar \hat{p}_k \\
\left[ \hat{p}_k , x_k \right] & = & -i \hbar \delta_{k,l}
\end{eqnarray} 

\noindent The angular ladder operators (\ref{eq17}) (\ref{eq18}) can be expressed in terms of the $\mathfrak{sl}(2) + \!\!\!\!\!\! \supset \mathbb{H}_n$ operators as 

\begin{eqnarray}
\label{lmeno}
\hat{L}^{-(n)}_k & \equiv & \frac{x_k}{\sqrt{\hat{J}_-^{(n)}}} \hat{L}^{(n)} - \frac{i}{\sqrt{\hat{J}_-^{(n)}}} \left( x_k \hat{J}_3^{(n)} - \hat{J}_-^{(n)} \hat{p}_k + i \hbar x_k  \right) \\
\label{lpiu}
\hat{L}^{+(n)}_k & \equiv & \hat{L}^{(n)} \frac{x_k}{\sqrt{\hat{J}_-^{(n)}}} + \frac{i}{\sqrt{\hat{J}_-^{(n)}}} \left( x_k \hat{J}_3^{(n)} - \hat{J}_-^{(n)} \hat{p}_k  \right) 
\end{eqnarray}

and they satisfy

\begin{eqnarray}
\label{lmenob}
(\hat{L}^{(n)})^2 \hat{L}^- & = & \hat{L}^- (\hat{L}^{(n)} - \hbar)^2 \\
\label{lpiub}
(\hat{L}^{(n)})^2 \hat{L}^+ & = & \hat{L}^+ (\hat{L}^{(n)} + \hbar)^2.
\end{eqnarray}

\section{The three dimensional system with spin and its third order integrals}
\subsection{The Hamiltonian and the integrals of the motion}
Because of the coalgebra symmetry we can state that the system $\hat{H}^{(3)}$ commutes with the Casimirs $\hat{\mathcal{C}}^{(2)}, \hat{\mathcal{C}}_{(2)} , \hat{\mathcal{C}}^{(3)}$ and furthermore with the  three components of 

\begin{eqnarray}
\label{cubicrungeX}
X_j &=& \frac{1}{\sqrt{2}} \left(\hat{L}^{+(3)}_j \hat{A}^{(3)} + \hat{A}^{(3)\dagger} \hat{L}^{-(3)}_j   \right)  ,\quad 1 \leq j \leq 3 \\
\label{cubicrungeY}
Y_j &=& \frac{1}{i \sqrt{2}} \left(\hat{L}^{+(3)}_j \hat{A}^{(3)} - \hat{A}^{(3)\dagger} \hat{L}^{-(3)}_j   \right),\quad 1 \leq j \leq 3
\end{eqnarray}

\noindent Equations (\ref{cubicrungeX}) , (\ref{cubicrungeY} ) are three dimensional analogs of (\ref{quadraticrungeX}), (\ref{quadraticrungeY}), however $X_j$, $Y_j$ are third order differential operators whereas $X$ and $Y$ are second order ones. 
The operators $\hat{H}^{(3)}$ and $X_j,Y_j $ depend linearly on the operator $\hat{L}^{(3)}$. Its square was defined in (\ref{Lalg}).  We need a realization of the operator $\hat{L}^{(3)}$ itself in order to obtain the three dimensional Hamiltonian  $\hat{H}^{(3)}$ and its integrals of motion $X_j,Y_j$. For a two dimensional system this is not an issue. In the 3-dimensional case $(\hat{L}^{(3)})^2$ turns out to be:

\begin{equation}
(\hat{L}^{(3)})^2 = \hat{\mathcal{C}}^{(3)} + \hbar^2 = \hat{L}_1^2 + \hat{L}_2^2 + \hat{L}_3^2 + \frac{\hbar^2}{4}
\end{equation}

As in the case of the Dirac equation, the square root of a sum of squared differential operators can be computed considering this operation on a space of anticommuting matrices, which in this case are the Pauli sigma matrices (\ref{paulisigma}). 

\noindent They allow us to introduce the following representation of $\hat{L}^{(3)}$

\begin{equation}
\hat{L}^{(3)} = \sqrt{\hat{\mathcal{C}}^{(3)} + \hbar^2 } = \sigma_i  \hat{L}_i + \frac{\hbar}{2}
\end{equation}
 
It can be verified by a direct calculation that $\hat{L}_i^{\pm}$ satisfy

\begin{equation}
\label{scalalineare}
[ \hat{L}^{(3)}, \hat{L}^{(\pm)}_i ] = \pm \hbar \hat{L}^{\pm}_i 
\end{equation}

This leads directly to the representation (\ref{eqa1.4}) of $\hat{H}^{(3)}$.
The relations $[X_j , H]=[Y_j , H]=0$ can be verified directly or by using the algebraic relations (\ref{eq37}, \ref{eq38}, \ref{lmenob}, \ref{lpiub}, \ref{scalalineare}).
The operators $\hat{L}_j^{+(3)}$ and $\hat{L}_j^{-(3)}$ are the three dimensional versions of the operators defined in (\ref{lmeno}), (\ref{lpiu}), and $\hat{A}^{(3)}$, $\hat{A}^{\dagger (3)}$of those defined  in (\ref{eq3.40}) (\ref{eq3.41}). 
Similarly as in the case of the Dirac equation, the calculation of the square root of a Casimir operator leads to the introduction of a spin term, in this case the spin-orbital interaction in 
(\ref{eqa1.4}). As mentioned in the introduction, a Hamiltonian of the form (\ref{eqa1.4}) was obtained in \cite{18} as part of a systematic search for superintegrable systems with second order integrals of motion. This lead to (\ref{eqa1.4}) with $\gamma = \frac{1}{2}$. Using the coalgebra symmetry approach we have obtained a more general result. The price of this generality is that the additional vector integrals of motion are third order operators $\vec{X}, \vec{Y}$ that reduce to  second order ones for $\gamma=\frac{1}{2}$ or $1$. 
 
Finally let us give an explicit representation for the constant of motion $\vec{X}$ which can be regarded as the generalization of the Laplace-Runge-Lenz vector for the hydrogen atom

\begin{eqnarray}
\vec{X}  & = & \frac{1}{2} ((\vec{L} \cdot \sigma) \vec{\mathcal{A}} + \vec{\mathcal{A}} (\vec{L} \cdot \sigma) ) + \frac{3 \hbar}{2} \vec{\mathcal{A}} \\
\hat{\vec{\mathcal{A}}} & \equiv & \frac{1}{2} \left( \hat{\vec{p}} \wedge  \hat{\vec{L}} - \hat{\vec{L}} \wedge \hat{\vec{p}}\right) + 2 \gamma \vec{p} \wedge \vec{S} + \frac{1}{2} \left( \vec{x} \hat{\mathcal{V}} + \hat{\mathcal{V}} \vec{x} \right) \\
\hat{\mathcal{V}} & \equiv & - \frac{\alpha}{r} + \frac{ 2 \gamma}{r^2} \vec{L} \cdot \vec{S} + \frac{\hbar^2 \gamma (2 \gamma + 1)}{2 r^2}
\end{eqnarray}

\subsection{The symmetry algebra associated with $\hat{H}^{(3)} $ }
The algebra generated by the set of constants of the motion for $\hat{H}_G^{(3)} $ defines a closed polynomial algebra under the operation of commutation. It is easy to compute this algebra if we consider the following fundamental identities:

\begin{eqnarray}
\label{fundcom}
\mathcal{J}_j &=& \L_j + \frac{\hbar}{2} \sigma_j \\
L^-_j L^+_k &=& -\frac{1}{2} \left( \mathcal{J}_j \mathcal{J}_k + \mathcal{J}_k \mathcal{J}_j \right) - i \epsilon_{jkl} \mathcal{J}_l (\vec{L} \cdot \vec{\sigma} + 2 \hbar) + \delta_{jk} ({\mathbf{\mathcal{J}}}^2 + \hbar (\vec{L} \cdot \vec{\sigma} + \frac{3}{2} \hbar)) \\  
L^+_j L^-_k &=& -\frac{1}{2} \left( \mathcal{J}_j \mathcal{J}_k + \mathcal{J}_k \mathcal{J}_j \right) + i \epsilon_{jkl} \mathcal{J}_l (\vec{L} \cdot \vec{\sigma} ) + \delta_{jk} ({\mathbf{\mathcal{J}}}^2 + \hbar (\vec{L} \cdot \vec{\sigma} + \frac{1}{2} \hbar)) 
\end{eqnarray}

\begin{eqnarray}
\hat{A}^\dagger (\hat{L}) \hat{A} ( \hat{L}) & = & (\hat{L} + \frac{\hbar}{2} + \hbar \gamma)^2 \hat{H}_G + \frac{\alpha^2}{2} \\
 \hat{A} (\hat{L} - \hbar ) \hat{A}^\dagger ( \hat{L} - \hbar ) & = & (\hat{L} - \frac{\hbar}{2} + \hbar \gamma)^2 \hat{H}_G + \frac{\alpha^2}{2} \\
\hat{\vec{L}}^{\pm} \cdot \hat{\vec{\mathcal{J}}} & = & 0
\label{funda}
\end{eqnarray}  

\noindent Taking into account (\ref{fundcom}) - (\ref{funda}) we obtain the following polynomial symmetry algebra

\begin{eqnarray}
\label{symmetryalgebraa}
\left[ \mathcal{J}_i  ,\mathcal{J}_j \right] & = & i \hbar \epsilon_{i j k} \mathcal{J}_k \\
\left[X_i  ,\mathcal{J}_j  \right] & = & i \hbar \epsilon_{i j k} X_k \\
\left[Y_i  ,\mathcal{J}_j  \right] & = & i \hbar \epsilon_{i j k} Y_k \\
\left[X_i , X_j \right] & =  & -i \hbar \epsilon_{i j k} \mathcal{J}_k  \mathcal{F}(H, L \cdot \sigma ) \\
\left[Y_i , Y_j \right] & = & -i \hbar \epsilon_{i j k} \mathcal{J}_k \mathcal{F}(H, L \cdot \sigma ) \\
\left[X_i , Y_j \right] & = & i \hbar ( L \cdot \sigma + \hbar (\gamma + \frac{1}{2}) ) (\mathcal{J}_i \mathcal{J}_j + \mathcal{J}_j \mathcal{J}_i) H + \delta_{i j} \mathcal{G}(H,L \cdot \sigma , {\mathbf{\mathcal{J}^2}}) \\
\left[ X_i , L \cdot \sigma \right] &=& -i \hbar Y_i \\
\label{symmetryalgebrab}
\left[ Y_i , L \cdot \sigma \right] &=& i \hbar X_i  
\end{eqnarray}

where 

\begin{equation}
\mathcal{F}(H, L \cdot \sigma) = \alpha^2 + H \left( 4 (L \cdot \sigma)^2 +\hbar (L \cdot \sigma) (6 \gamma + 5) + 2 \hbar^2 (\gamma+1)^2 \right)
\end{equation}

\begin{eqnarray}
\mathcal{G} &=& \frac{-i \hbar}{2} ( 2 \alpha^2 (L \cdot \sigma + \hbar) + H ( 4 (L \cdot \sigma) ({\mathbf{\mathcal{J}}}^2  + (L \cdot \sigma)^2 ) +2\hbar ({\mathbf{\mathcal{J}}}^2(1+2 \gamma) +4(L \cdot \sigma)^2 (1+ \gamma) ) + \nonumber \\
&& + 4 \hbar^2 (L \cdot \sigma) (1+\gamma)(2+ \gamma) + \hbar^3 (3+6\gamma + 4 \gamma^2) ) 
\end{eqnarray}

\noindent All commutators not shown above vanish. The basis elements of the algebra are $\{H,\mathcal{J}_i, X_i, Y_i, (\vec{\sigma} , \vec{L}) , 1 \}$ and the right hand sides are at most fourth order polynomials in the basis elements. 

\section{Exact bound states solutions of the Schr\"odinger-Pauli equation}
Let us conclude the analysis of this Hamiltonian system by evaluating explicitly its eigenfunctions and its spectrum for bound states. We construct the wavefunction as a complete set of commutative operators

\begin{eqnarray}
H \psi(r, \theta, \phi)_{q,n,j,k} & = & E \psi(r, \theta, \phi)_{q,n,j,k} \\ 
\label{spher1}
\hat{{\mathbf \mathcal{J}}}^2 \Omega(\theta, \phi)_{q, j, k} & = & j (j+1) \Omega(\theta, \phi)_{q, j, k} 
\\
\hat{\mathcal{J}}_{3} \Omega(\theta, \phi)_{q,j, k} & = & k \Omega(\theta, \phi)_{q,j, k} 
\\
\vec{L} \cdot \vec{S}  \Omega(\theta, \phi)_{q, j, k}&=& \frac{q}{2}  \Omega(\theta, \phi)_{q, j, k}; q=
\begin{cases}
l \\
-l -1
\end{cases} \\
\hat{L}^2 \Omega(\theta, \phi)_{q, j, k} & = & q(q+1) \Omega(\theta, \phi)_{q, j, k}
\end{eqnarray}

\begin{eqnarray}
\psi(r, \theta, \phi)_{q,n,j,k} &=& \rho(r)_{q,n,j} \Omega(\theta, \phi)_{q,j, k} \\
\Omega(\theta, \phi)_{l,j, k} & = & \frac{1}{\sqrt{2 j}} \left(
\begin{array}{c} \sqrt{j+ k} Y_{j-\frac{1}{2}, k - \frac{1}{2}} (\theta, \phi) \\
 \sqrt{j - k} Y_{j-\frac{1}{2}, k + \frac{1}{2}} (\theta, \phi)
\end{array}
\right)
\\
\label{spher2}
\Omega(\theta, \phi)_{-l-1,j, k} & = & \frac{1}{\sqrt{2 j + 2}} \left(
\begin{array}{c} \sqrt{j - k +1 } Y_{j+\frac{1}{2}, k - \frac{1}{2}} (\theta, \phi) \\
 \sqrt{j + k +1} Y_{j + \frac{1}{2}, k + \frac{1}{2}} (\theta, \phi)
\end{array}
\right)
\end{eqnarray}

and the functions $Y_{l,m}(\theta, \phi)$ are the usual spherical harmonic functions:
\begin{eqnarray}
\hat{L}^2 Y_{l,m}(\theta, \phi) = l(l+1) Y_{l,m}(\theta, \phi) \\
\hat{L}_3 Y_{l,m}(\theta, \phi) = m Y_{l,m}(\theta, \phi).
\end{eqnarray}

In view of (\ref{spher1}) - (\ref{spher2}) we can reduce the 3-dimensional Hamiltonian operator $\hat{H}$ to the following radial one:

\begin{eqnarray}
\hat{\mathcal{H}} = < \Omega(\theta, \phi)_{q, j, k}| \hat{H}|  \Omega(\theta, \phi)_{q,j, k}> = 
\\
= \begin{cases}
q= l \rightarrow
-\frac{\hbar^2}{2} \left( \partial_r^2 + \frac{2}{r} \partial_r + \frac{(l+\gamma)(l+\gamma +1)}{r^2}\right) - \frac{\alpha}{r}
\\
q=-l-1 \rightarrow
-\frac{\hbar^2}{2} \left( \partial_r^2 + \frac{2}{r} \partial_r + \frac{(l - \gamma)(l - \gamma +1)}{r^2}\right) - \frac{\alpha}{r}.
\end{cases}
\end{eqnarray}

It is straightforward to get the explicit expression for the bound state eigenfunctions of $\hat{\mathcal{H}}$ 

\begin{eqnarray}
\rho_{l , n, j} \propto r^{j + \gamma -\frac{1}{2}} e^{-\frac{\alpha}{\hbar^2 (n + \gamma + j + \frac{1}{2})}r} L_n^{2j +2 \gamma}( \frac{2 \alpha r}{\hbar^2 (n + \gamma + j +\frac{1}{2})})
\\
\rho_{-l-1 , n, j} \propto r^{j - \gamma +\frac{1}{2}} e^{-\frac{\alpha}{\hbar^2 (n - \gamma + j + \frac{3}{2})}r} L_n^{2j - 2 \gamma + 2} ( \frac{2 \alpha r}{\hbar^2 (n - \gamma + j +\frac{3}{2})})
\end{eqnarray}

\noindent $L_n^k(x)$ are Laguerre polynomials. The bound state normalization condition requires for $\rho_{l,n,j}$ that $j > -\gamma -1$ and for $\rho_{-l-1,n,j}$ that $j > \gamma -2$. Finally we obtain for the bound spectrum 

\begin{eqnarray}
\hat{\mathcal{H}} \rho_{l ,n,j} (r) = -\frac{ \alpha^2}{2 \hbar^2 (n + j + \gamma +\frac{1}{2})^2 } \rho_{l ,n,j} (r) \\
\hat{\mathcal{H}} \rho_{-l-1 ,n,j} (r) = -\frac{ \alpha^2}{2 \hbar^2 (n + j - \gamma +\frac{1}{2})^2 } \rho_{-l-1 ,n,j} (r)
\end{eqnarray}

\section{Conclusions}
The main physical result of this paper is the new superintegrable system with spin described by the Hamiltonian (\ref{eqa1.4}) and the polynomial algebra of integrals of motion (\ref{symmetryalgebraa})-(\ref{symmetryalgebrab}). The corresponding Schr\"odinger Pauli equation is solved in Section 5 where we give the bound state energies and wave functions. The radial parts are Laguerre polynomials times factors ensuring appropriate behaviour for $r \rightarrow 0$ and $r \rightarrow \infty$. The angular parts are expressed in terms of spherical spinors. 
A special case of the Hamiltonian (\ref{eqa1.4})with $\gamma = \frac{1}{2}$ was obtained in \cite{18} as part of a systematic search for superintegrable system of the form (\ref{eqi2}) with integrals of motion of degree at most 2 in the momenta \cite{17}, \cite{18}.
Another special case of (\ref{eqa1.4}) with $\gamma = 1$ is implicit in \cite{17}, \cite{18}. It is obtained from the spinless Coulomb Hamiltonian $H = -\frac{1}{2} \Delta + \frac{\alpha}{r}$ in $\mathbb{E}_3$ by a gauge transformation (given in \cite{17}) that transforms the total angular momentum $\vec{L}$ into a new integral of motion which depends linearly on the total angular momentum $\vec{\mathcal{J}}$ and appropriately transforms the Laplace-Runge-Lenz vector.
The power of the coalgebra approach is that it leads to a general independent parameter $\gamma$ in the spin orbital potential and that it leads directly to the third order integrals of motion $\vec{X}$ , $\vec{Y}$. A systematic search for systems with third order integrals would be very difficult. The explicit form of the integrals $\vec{X}$, $\vec{Y}$ is actually quite simple , however it is not linear in the Pauli matrices $\sigma_i$. The quadratic terms can be eliminated, but that leads to quite complicated expressions. For instance we obtain

\begin{eqnarray}
&& \vec{X} =  \vec{x} \left(-\frac{ \alpha}{r} + \frac{\hbar^2 \gamma (\gamma + 1)}{r^2} + (\vec{L} \cdot \vec{\sigma}) {\mathbf{p}}^2  + \hbar (1 + 2 \gamma) {\mathbf{p}}^2  + \frac{2 i \hbar^2 \gamma}{r^2} (\vec{x} \cdot \vec{p}) - \frac{\hbar \gamma}{r^2} (\vec{x} \cdot \vec{p})^2\right) \nonumber \\
&& + \left( -(\vec{L} \cdot \vec{\sigma}) (\vec{x} \cdot \vec{p}) + i \hbar (\vec{L} \cdot \vec{\sigma})  -i \hbar^2 \gamma -\hbar (1 + \gamma) (\vec{x} \cdot \vec{p}) +i \hbar^2 (1+ \gamma) + \left(- \frac{\alpha}{r} \! + \! \frac{\hbar^2 \gamma^2}{r^2} \right) \!\! (\vec{\sigma} \cdot \vec{L}) \right) \vec{p} \nonumber \\
&& + (\vec{x} \wedge \vec{\sigma}) \left( \frac{i \hbar}{2} {\mathbf{p}}^2   + \frac{i \hbar}{2} \frac{\alpha}{r}  -  \frac{\hbar^2 \gamma}{2 r^2} (\vec{x} \cdot \vec{p})  \right) + \left(\frac{\hbar^2 (1 + 2 \gamma)}{2}  + \frac{i \hbar}{2} (\vec{x} \cdot \vec{p}) \right) (\vec{p} \wedge \vec{\sigma}) \nonumber
\end{eqnarray}

\noindent and $\vec{Y}$ is similar. 
In the past systematic searches for second order superintegrable systems were conducted for purely scalar potentials $V_0(\vec{r})$ in $\mathbb{E}_2$ and $\mathbb{E}_3$ \cite{26},\cite{27},\cite{28} and also in more general conformally flat spaces \cite{29}.  Searches for third order supereintegrable systems in $\mathbb{E}_2$, allowing separation of variables were also succesful \cite{30},\cite{31},\cite{32}, but were considerably more difficult. For particles with spin a systematic approach to searching for higher order integrable and superintegrable systems is a prohibitive task. Hence the development of other techniques is imperative, in particular those involving coalgebra symmetry.  
Further applications and generalizations of the coalgebra techniques are in progress, for instance to obtain  suprintegrable systems in $n$ dimensions and systems involving higher spins, or two particle with nonzero spin.

 \section{Acknowledgments}
We thank I. Yurdusen and A. Nikitin for interesting discussion. D.R. acknowledges a postdoctoral fellowship from the laboratory of Mathematical Physics of the CRM, O.G. an undergraduate summer (ISM-CRM) fellowship. The research of P.W. was partially supported by a research grant from NSERC of Canada.     


\section*{References}
\begin{thebibliography}{99}


\bibitem{1}
Fock V A 1935 Zur Theorie des Wasserstoffatoms Z. Phys. 98 145–54

\bibitem{2}
Bargmann V 1936 Zur Theorie des Wasserstoffatoms Z. Phys. 99 576–82

\bibitem{3}
Jauch J and Hill E 1940 On the problem of degeneracy in quantum mechanics Phys. Rev. 57 641–5

\bibitem{4}
W Miller Jr, Post S.and Winternitz P 2013 Classical and quantum superintegrability with applications J. Phys. A: Math. Theor. 46 423001 

\bibitem{5}
Tempesta P, Turbiner V and Winternitz P 2001 Exact solvability of superintegrable systems J. Math.Phys. 42 419–36

\bibitem{6}
Daboul J, Slodowy P and Daboul C 1993 The hydrogen algebra as centerless twisted Kac–Moody algebra Phys. Lett. B 317 321–8

\bibitem{7}
Daskaloyannis C 2001 Quadratic Poisson algebras of two-dimensional classical superintegrable systems and quadratic associative algebras of quantum superintegrable systems J. Math. Phys. 42 1100–19

\bibitem{8}
Pronko G P and Stroganov Y G 1977 New example of quantum mechanical problem with hidden symmetry Sov. Phys.JETP 45 1075–77

\bibitem{9}
Pronko G P 2007 Quantum superintegrable systems for arbitrary spin J. Phys. A: Math. Theor. 40 1333–36

\bibitem{10} 
D'Hoker E and Vinet L 1984 Supersymmetry of the Pauli equation in the presence of a magnetic monopole  Phys. Lett. B, 1984  137  72–76

\bibitem{11}
D'Hoker E and Vinet L 1984  Dynamical supersymmetry of the magnetic monopole and the 1/r 2-potential – Comm. Math.Phys. 97, 391-427 (1985)

\bibitem{12}
Nikitin A G 2013   Superintegrable systems with spin invariant with respect to the rotation group
 J. Phys. A: Math. Theor. 46 265204 


\bibitem{14}
Nikitin A G 2012 New exactly solvable systems with Fock symmetry J. Phys. A: Math. Theor. 45 485204 

\bibitem{13}
Nikitin A G 2012 Matrix superpotentials and superintegrable systems for arbitrary spin J. Phys. A: Math. Theor. 45 225205

\bibitem{15}
Nikitin A G and Karadzhov Y  2011 Enhanced classification of matrix superpotentials  J. Phys. A: Math. Theor. 44 445202 

\bibitem{16}
Winternitz P and Yurdusen I 2006 Integrable and superintegrable systems with spin J. Math. Phys. 47 103509

\bibitem{17}
Winternitz P and Yurdusen I 2009 Integrable and superintegrable systems with spin in three-dimensional Euclidean space J. Phys. A: Math. Theor. 42 385203 

\bibitem{18}
Desilets J-F, Winternitz P and Yurdusen I 2012 Superintegrable systems with spin and second-order integrals of motion J. Phys. A: Math. Theor. 45 475201 

\bibitem{19}
Johnson M.A. and Lippman B.A.. Relativistic Kepler problem, Phys Rev. 78, 329 (1950)

\bibitem{20}
Katsura H and Aoki H 2006
Exact supersymmetry in the relativistic hydrogen atom in general dimensions. Supercharge and the generalized Johnson-Lippmann operator J. Math. Phys. 47 , 032301  

\bibitem{21}
A Ballesteros, A Blasco, F J Herranz, F Musso and O Ragnisco 2009 (Super)integrability from coalgebra symmetry: Formalism and applications . J. Phys.: Conf. Ser. 175 012004

\bibitem{22}
Riglioni D 2013 Classical and quantum higher order superintegrable systems from coalgebra symmetry J. Phys. A: Math. Theor. 46 265207

\bibitem{23}
Tremblay F, Turbiner V A and Winternitz P 2009 An infinite family of solvable and integrable quantum systems on a plane J. Phys. A: Math. Theor. 42 24200

\bibitem{24}
Tremblay F, Turbiner V A and Winternitz P 2010 Periodic orbits for an infinite family of classical superintegrable systems J. Phys. A: Math. Theor. 43 015202

\bibitem{25}
Post S. and Winternitz P. 2010 
An infinite family of superintegrable deformations of the Coulomb potential J. Phys. A: Math. Theor. 43 222001

\bibitem{26}
Fris I, Mandrosov V, Smorodinsky Ja A, Uhlir M and Winternitz P 1965 On higher symmetries in quantum mechanics Phys. Lett. 16 354–6

\bibitem{27}
Winternitz P, Smorodinsky J, Uhlir˘ M and Fris˘ I 1966 Symmetry groups in classical and quantum mechanics Yad. Fiz. 4 625–35 [Sov. J. Nucl. Phys. 4 444–50 1967(Engl. transl.)]

\bibitem{28}
Makarov A A, Smorodinsky Ja A, Valiev Kh. and Winternitz P 1967 A systematic search for nonrelativistic systems with dynamical symmetries Il Nuovo Cimento A 52 1061–84

\bibitem{29}
Kalnins E G, Kress J M and Miller W Jr 2005 Second order superintegrable systems in conformally flat spaces: I,...,V J. Math. Phys. 46 053509, 053510, 103507, 47 043514, 093501

\bibitem{30}
Gravel S 2004 Hamiltonians separable in Cartesian coordinates and third-order integrals of motion J. Math. Phys. 45 1003–19

\bibitem{31}
Gravel S and Winternitz P 2002 Superintegrability with third-order integrals in quantum and classical mechanics J. Math. Phys. 46 5902

\bibitem{32}
Tremblay F and Winternitz P 2010 Third order superintegrable systems separating in polar coordinates J. Phys. A: Math. Theor. 43 175206


\end{thebibliography}
\end{document}